*Brief report*

# **Characterization of Methicillin-resistant *Staphylococcus aureus* Isolates from Fitness Centers in Memphis Metropolitan Area, USA**

Nabanita Mukherjee [a], Irshad M. Sulaiman [b], Pratik Banerjee [a*]

[a] Division of Epidemiology, Biostatistics, and Environmental Health, School of Public Health, The University of Memphis, Memphis, Tennessee 38152, USA

[b] Southeast Regional Laboratory, Microbiological Sciences Branch, U. S. Food and Drug Administration, Atlanta, Georgia 30309, USA

** Corresponding Author:*
Pratik Banerjee, Ph.D.
Division of Epidemiology, Biostatistics, and Environmental Health
School of Public Health
The University of Memphis
338 Robison Hall, Memphis, Tennessee 38152, USA.
Tel: +001-901-678-4443
Fax: +001-901-678-4504
E-mail: pbnerjee@memphis.edu

**Abstract**

Indoor skin-contact surfaces of public fitness centers may serve as reservoirs of potential human transmission of methicillin-resistant *Staphylococcus aureus* (MRSA). We found a high prevalence of multi-drug resistant (MDR)-MRSA of CC59 lineage harboring a variety of extracellular toxin genes from surface swab samples collected from inanimate surfaces of fitness centers in Memphis metropolitan area, USA. Our findings underscore the role of inanimate surfaces as potential sources of transmission of MDR-MRSA strains with considerable genetic diversity.

Community-associated methicillin-resistant *Staphylococcus aureus* (CA-MRSA), contagion with varied severity ranging from superficial skin and soft tissue infections to life-threatening conditions, including endocarditis, bacteremia, and necrotizing pneumonia, has emerged as the major cause of human staphylococcal infections.[1, 2] Recent literature reported the diversity in the clonal and molecular characteristics of CA-MRSA, making it almost indistinguishable from its healthcare-associated counterpart (HA-MRSA).[3] Inanimate surfaces may serve as environmental reservoirs for CA-MRSA transmission; and for public fitness and sports facilities where skin contact is common, this amounts to be a matter of grave concern.[1] However, the information on the molecular characteristics of the environmental MRSA isolates from fitness centers and sports facilities is limited, and therefore warrants more investigation. In a recent metagenomics-based study reporting the overall bacterial diversity of the common fitness center surfaces revealed an abundance of operation taxonomic units (OTUs) attributable to *Staphylococcus* species.[4] In this brief communication, we report the prevalence, multi-drug resistance, and molecular characteristics of the MRSA isolates from fitness centers in Memphis metropolitan area in Tennessee.

## METHODS

The sample collection procedures have been described in a previous study.[4] Briefly, a total of 32 samples were collected from the skin-contact surfaces on exercise equipment (nautilus machine, stationary bike, treadmill, elliptical machine, power strider, and leg press), dumbbell, toilet handles, and handrails on stairs from four membership-based fitness centers around the Memphis metropolitan area. The presumptive *S. aureus* colonies were isolated and identified by plating on BBL™ CHROMagar™ MRSA plates (BDCM) (BD Diagnostics, USA) followed by sub-culturing on Mannitol Salt agar (MSA) (Thermo Scientific™ Remel™ MSA, USA), and by coagulase and catalase tests. Further, the molecular identification of the cultures was done by amplifying staphylococcal 16S rRNA gene, as well as *mecA*, *femA*, and *femB* genes associated with the methicillin resistance property of *S. aureus* (details of the methods are provided in the Supplementary Information). Antimicrobial susceptibility testing was performed using the Sensititre® microbroth dilution system (Trek Diagnostic Systems Inc., Cleveland, USA) following the manufacturer's instructions and the susceptibility was determined according to Clinical and Laboratory Standards Institute guidelines. MRSA positive samples were characterized by PFGE; MLST; SCC*mec*, and *ccr* typing. Further, the presence of genes encoding for enterotoxin (*sea*, *seb*, *sec*, *sed* and *see*), exfoliative toxin (*eta* and *etb*), toxic shock syndrome toxin 1 (*tst*), and Panton-Valentine leukocidin cytotoxin (*pvl*) were evaluated according to the methods described previously (Supplementary Information).

## RESULTS AND DISCUSSION

Twenty-nine samples (90.6%) yielded presumptive positive colonies for *Staphylococcus* spp. Among these, 12 (37.5%) isolates were positive for MRSA as determined by the presence of *mecA*, *femA*, and *femB* genes. All MRSA isolates exhibited resistance characteristics to multiple (7 to12) antibiotics. The MLST analysis revealed that all isolates belonged to clonal complex 59 (CC59) or sequence type 59 (ST59), a distinct CA-MRSA genotype reported in other countries.[5] The distribution of enterotoxin genes among the twelve isolates showed the following profile: *sec* (12 isolates), *sea* (9 isolates), *sed* (9 isolates), *see* (9 isolates), and *seb* (6 isolates). Toxic shock syndrome toxin-1 gene (*tst*) was detected in all twelve the isolates, however, none of the isolates carried the *pvl* gene. All isolates were *eta* positive but *etb* negative. The PFGE analysis revealed ten distinct pulse-field types (PFT) with three of the isolates (S3, S10, and S14) shared identical PFT (Figure 1). Among these identical PFT isolates, S10 and S14 were recovered from toilet handle and dumbbell, respectively of the same fitness center while S3 was isolated from a stationary bike of another facility.

In our study, all MRSA isolates had a common lineage - CC59/ST59 with multifarious genotypic background. However, unlike typical CA-MRSA, the isolates were found to harbor a type II SCC*mec* element and did not carry *pvl* genes. A significantly heterogeneous global epidemiology of MRSA clones (specially CA-MRSA) is apparent from several recent reports.[3] Likewise, the CC59 clones isolated in the current study showed a unique combination of genotypic traits. Previous reports from various parts of the world have shown significant genetic diversity of CC59 strains.[5, 6] For example, the "Taiwan clone" was reported to carry the type $V_T$ SCC*mec* element and *pvl* genes, while the "Asian-Pacific clone" usually carried a type IV SCC*mec* element but no *pvl* genes.[6] In Western Australia, at least six different groups of CC59 strains had been isolated.[5]

The current study is the first report of the prevalence of CC59 MRSA in indoor environments in the US. Most of the CC59 isolates reported from other countries were human isolates (e.g., isolated from skins, soft tissues, nasal swabs, etc.). We believe the CC59 clones found in our study may have been transferred from humans to the commonly used equipment surfaces of fitness center as a result of contact with human skin. Another important finding of the current study is that all isolates showed resistance to clindamycin and tetracycline. In addition, most of the strains showed resistance to multiple other antibiotics. Half of the isolates (6 out of 12) recovered from this study were found to be resistant to linezolid which is one of the few antibiotics reported to be effective against MRSA.[7] To the best of our knowledge; the present study is the first report of isolation and prevalence of linezolid-resistant MRSA strains from Memphis area. Previously a large hospital outbreak of linezolid-resistant enterococci (LRE) infection has been reported from this area.[8] The levels of antibiotic resistance were shown to vary significantly in CC59 MRSA, for example, CC59 strains from Taiwan showed much higher multi-drug resistance compared to a CC59 MRSA-type strain USA1000.[6]

Healthcare-associated infection (HAI) caused by MRSA has been a long-standing challenge in Memphis area as reported by the Tennessee Department of Health (TDH). The TDH HAI database reveals that several healthcare facilities in this area frequently report a standardized infection ratio (SIR) of 2.0 or more which is above the national guideline of 1.0.[9] Our study reveals a high prevalence of MDR-MRSA of CC59 lineage consisting of a variety of extracellular toxins in a relatively small catchment area (Memphis metropolitan) and limited sample size. Recently, the presence of high-level multi-drug resistant (MDR)-MRSA was also reported from recreational beaches and in built environment in the US.[2] However, the emergence of CC59 MRSA in the US may be a matter of concern as clones belonging to this sequence types are known to colonize very efficiently in the nasal cavities of children.[10] Moreover, acquisition of multidrug resistance mechanism in CC59 is significantly different from many other MRSA strains and these strains are reported to harbor more resistance genes than other MRSA types. [5, 10]

In this study, we report the microbiological results from environmental samples, and therefore these results alone cannot be associated with the colonization or infections of the human population (such as users of the gym). Nonetheless, this highlights the role of inanimate environmental reservoirs as sources for potential human transmission of MDR-MRSA. The evolution and genetic variability of CC59 strains like the ones detected in our study remain to be understood. Further research on the possible modes of transmission, distribution, and carriage of theses MRSA strains in human host along with public health awareness around increased standards for maintaining hygiene in public facilities to reduce the risk of infection are warranted.

## Acknowledgments

This work was partly supported by start-up funds from University of Memphis, Memphis Research Consortium, and a grant from the U.S. Food & Drug Administration (1U54FD004330-01) to P.B. We thank Dr. John Dunn, Sheri Roberts, Linda Thomas, and Jeannette Dill (Tennessee Department of Health) for PFGE analysis, Emily Jacobs (FDA) for technical assistance, and Satish Kedia for critically reading the manuscript.

## References

1. Cohen PR. The skin in the gym: a comprehensive review of the cutaneous manifestations of community-acquired methicillin-resistant *Staphylococcus aureus* infection in athletes. Clin Dermatol 2008;26:16-26.

2. Roberts MC, Soge OO, No D. Comparison of Multi-Drug Resistant Environmental Methicillin-Resistant *Staphylococcus aureus* Isolated from Recreational Beaches and High Touch Surfaces in Built Environments. Front Microbiol 2013;4:74.

3. Chua K, Laurent F, Coombs G, Grayson ML, Howden BP. Not community-associated methicillin-resistant *Staphylococcus aureus* (CA-MRSA)! A clinician's guide to community MRSA - its evolving antimicrobial resistance and implications for therapy. Clin Infect Dis 2011;52:99-114.

4. Mukherjee N, Dowd SE, Wise A, Kedia S, Vohra V, Banerjee P. Diversity of bacterial communities of fitness center surfaces in a U.S. metropolitan area. Int J Environ Res Public Health 2014;11:12544-61.

5. Coombs GW, Monecke S, Ehricht R, et al. Differentiation of clonal complex 59 community-associated methicillin-resistant *Staphylococcus aureus* in Western Australia. Antimicrob Agents Chemother 2010;54:1914-21.

6. Chen CJ, Unger C, Hoffmann W, Lindsay JA, Huang YC, Gotz F. Characterization and comparison of 2 distinct epidemic community-associated methicillin-resistant *Staphylococcus aureus* clones of ST59 lineage. PLoS One 2013;8:e63210.

7. Gu B, Kelesidis T, Tsiodras S, Hindler J, Humphries RM. The emerging problem of linezolid-resistant *Staphylococcus*. J Antimicrob Chemother 2013;68:4-11.

8. Kainer MA, Devasia RA, Jones TF, et al. Response to emerging infection leading to outbreak of linezolid-resistant enterococci. Emerg Infect Dis 2007;13:1024-30.

9. TDH. Tennessee Reports on Healthcare Associated Infections - URL: https://tn.gov/health/article/cedep-reports#sthash.JbLSPEI9.dpuf (accessed on June 17, 2016). 2015.

10. Aung MS, Zi H, Nwe KM, et al. Drug resistance and genetic characteristics of clinical isolates of staphylococci in Myanmar: high prevalence of PVL among methicillin-susceptible *Staphylococcus aureus* belonging to various sequence types. New Microbes New Infect 2016;10:58-65.

**Figure and Figure Legend.**

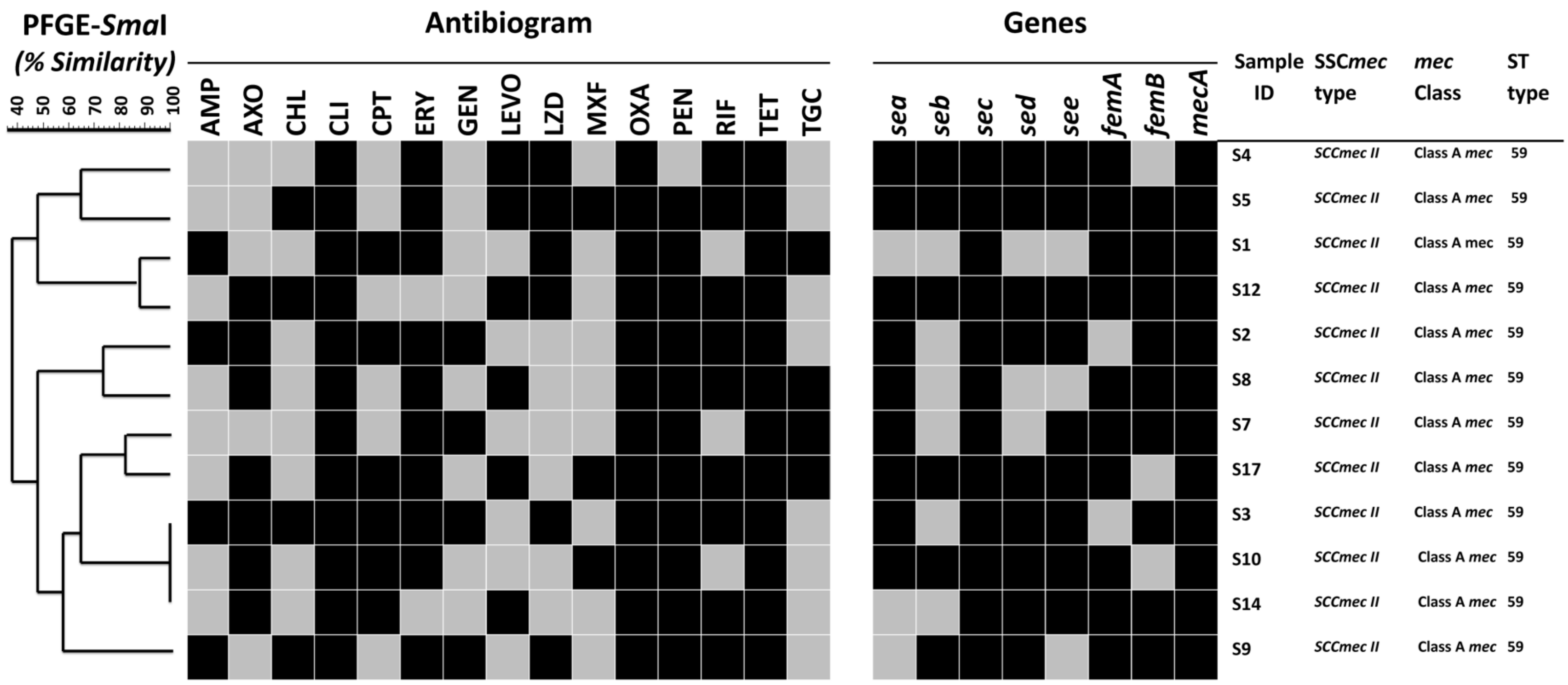

**Figure 1.** Dendogram showing the comparison of *Sma*I pulsed-field gel electrophoresis (PFGE) patterns along with genotypes of methicillin-resistant *Staphylococcus aureus* (MRSA) isolates from equipment surface of fitness centers. For PFGE, levels of similarity were determined using the Dice similarity coefficient and unweighted pair group method (UPGMA). All MRSA isolates were *eta,* and TSST-1 positive, *pvl* negative, carried type II SSC*mec*, and belonged to clonal lineage ST59 (described in text). All isolates were resistant to oxacillin. Most of the isolates were resistant to penicillin. The resistance profiles of the isolates are depicted in the antibiogram, where: AMP, ampicillin; AXO, ceftriaxone; CHL, chloramphenicol; CLI, clindamycin; CPT, ceftaroline; ERY, erythromycin; GEN, gentamicin; LEVO, levofloxacin; LZD, linezolid; MXF, moxifloxacin; OXA, oxacillin; PEN, penicillin; RIF, rifampin; TET, tetracycline; TGC, tigecycline. Color keys: Positive (■) ; negative (■).

## ***Supplementary Information***

**Characterization of Methicillin-resistant *Staphylococcus aureus* Isolates from Fitness Centers in Memphis Metropolitan Area, USA**

**Methods Details for Molecular Characterizations**

*SSC*mec *genotyping.* Multiplex PCR was performed for SCC*mec* and *ccr* typing as described previously. [1-3] Briefly, genomic DNA were extracted from 12 isolates using MO BIO UltraClean® Microbial DNA Isolation Kit (MO BIO Lab, CA) following the manufacturer's recommendations. Eight pairs of primers were used for SCC*mec* types I, II, III, IVa, IVb, IVc, IVd, and V, 2 pairs of primers for the *mec* gene, 4 primers for types 1, 2, or 3 *ccr*, and 1 primer pair for type 5 *ccr*. The details of primers used for SCC*mec* and *ccr* typing are described in Table S1.

*Identification of toxin genes by PCR.* A multiplex PCR was performed to identify the presence *Staphylococcal* toxin genes (*sea, seb, sec, sed, see, eta, etb,* and *tst*) as described previously.[1, 4] The details about the primers are described in Table S2. Additionally, another PCR reaction containing lukFS-PV primers (Table S2) was performed for identifying the Panton-Valentine leukocidin (PVL) toxin gene following the protocol described earlier.[1]

*Pulsed-field gel electrophoresis (PFGE) and multi-locus sequence typing (MLST).* Isolates were subjected to PFGE and MLST as described previously. [5-7]

**Table S1: Details of Primers used for SCC*mec* and *ccr* typing.**

| Primer Name | Primer sequence (5'-3') | Specificity | Reference |
|---|---|---|---|
| Type I-F | GCTTTAAAGAGTGTCGTTACAGG | SCC*mec* I | [3] |
| Type I-R | GTTCTCTCATAGTATGACGTCC | | |
| Type II -F | CGTTGAAGATGATGAAGCG | SCC*mec* II | [3] |
| Type II-R | CGAAATCAATGGTTAATGGACC | | |
| Type III-F | CCATATTGTGTACGATGCG | SCC*mec* III | [3] |
| Type III-R | CCTTAGTTGTCGTAACAGATCG | | |
| Type IVa-F | GCCTTATTCGAAGAAACCG | SCC*mec* IVa | [3] |
| Type IVa-R | CTACTCTTCTGAAAAGCGTCG | | |
| Type IVb-F | TCTGGAATTACTTCAGCTGC | SCC*mec* IVb | [3] |
| Type IVb-R | AAACAATATTGCTCTCCCTC | | |
| Type IVc-F | ACAATATTTGTATTATCGGAGAGC | SCC*mec* IVc | [3] |
| Type IVc-R | TTGGTATGAGGTATTGCTGG | | |
| Type IVd-F5 | CTCAAAATACGGACCCCAATACA | SCC*mec* IVd | [3] |
| Type IVd-R6 | TGCTCCAGTAATTGCTAAAG | | |
| Type V-F | GAACATTGTTACTTAAATGAGCG | SCC*mec* V | [3] |
| Type V-R | TGAAAGTTGTACCCTTGACACC | | |
| mecI-F | CCCTTTTTATACAATCTCGTT | Class A *mec* | [3] |
| mecI-R | ATATCATCTGCAGAATGGG | | |
| IS1272-F | TATTTTTGGGTTTCACTCGG | Class B *mec* | [3] |
| mecR1-R | CTCCACGTTAATTCCATTAATACC | | |
| ccrAB-β2 | ATTGCCTTGATAATAGCCITCT | *ccr* Types 1 through 5 | [2], [3] [3] |
| ccrAB-α2 | AACCTATATCATCAATCAGTACGT | | |
| ccrAB-α3 | TAAAGGCATCAATGCACAAACACT | | |
| ccrAB-α4 | AGCTCAAAAGCAAGCAATAGAAT | | |
| ccrC-F | ATGAATTCAAAGAGCATGGC | | |
| ccrC-R | GATTTAGAATTGTCGTGATTGC | | |

**Table S2: Details of Primers used for *Staphylococcal* toxin genes.**

| Gene name | Primer | Primer sequence (5'-3') | Reference |
|---|---|---|---|
| *sea* | GSEAR-1 | GGTTATCAATGTGCGGGTGG | [4] |
| | GSEAR-2 | CGGCACTTTTTTCTCTTCGG | |
| *seb* | GSEBR-1 | GTATGGTGGTGTAACTGAGC | [4] |
| | GSEBR-2 | CCAAATAGTGACGAGTTAGG | |
| *sec* | GSECR-1 | AGATGAAGTAGTTGATGTGTATGG | [4] |
| | GSECR-2 | CACACTTTTAGAATCAACCG | |
| *sed* | GSEDR-1 | CCAATAATAGGAGAAAATAAAAG | [4] |
| | GSEDR-2 | ATTGGTATTTTTTTTCGTTC | |
| *see* | GSEER-1 | AGGTTTTTTCACAGGTCATCC | [4] |
| | GSEER-1 | CTTTTTTTTCTTCGGTCAATC | |
| *eta* | GETAR-1 | GCAGGTGTTGATTTAGCATT | [4] |
| | GETAR-2 | AGATGTCCCTATTTTTGCTG | |
| *etb* | GETBR-1 | ACAAGCAAAAGAATACAGCG | [4] |
| | GETBR-2 | GTTTTTGGCTGCTTCTCTTG | |
| *tst* | GTSSTR-1 | ACCCCTGTTCCCTTATCATC | [4] |
| | GTSSTR-2 | TTTTCAGTATTTGTAACGCC | |
| *femA* | GFEMAR-1 | AAAAAAGCACATAACAAGCG | [4] |
| | GFEMAR-2 | GATAAAGAAGAAACCAGCAG | |
| *lukFS-PV* | F | TTACACAGTTAAAGAA | [1] |
| | R | AATGCAATTGATG | |

**References (*Supplementary Information*):**

1. Champion A, Goodwin TA, Brolinson P, Werre SR, Prater M, Inzana TJ. Prevalence and characterization of methicillin-resistant *Staphylococcus aureus* isolates from healthy university student athletes. Ann Clin Microbiol Antimicrob 2014;13:33.

2. Ito T, Katayama Y, Asada K, et al. Structural comparison of three types of staphylococcal cassette chromosome mec integrated in the chromosome in methicillin-resistant *Staphylococcus aureus*. Antimicrob Agents Chemother 2001;45:1323-36.

3. Zhang K, McClure JA, Elsayed S, Louie T, Conly JM. Novel multiplex PCR assay for characterization and concomitant subtyping of staphylococcal cassette chromosome *mec* types I to V in methicillin-resistant *Staphylococcus aureus*. J Clin Microbiol 2005;43:5026-33.

4. Mehrotra M, Wang G, Johnson WM. Multiplex PCR for detection of genes for *Staphylococcus aureus* enterotoxins, exfoliative toxins, toxic shock syndrome toxin 1, and methicillin resistance. J Clin Microbiol 2000;38:1032-5.

5. Jackson CR, Davis JA, Barrett JB. Prevalence and Characterization of Methicillin-Resistant Staphylococcus aureus Isolates from Retail Meat and Humans in Georgia. Journal of Clinical Microbiology 2013;51:1199-1207.

6. Enright MC, Day NP, Davies CE, Peacock SJ, Spratt BG. Multilocus sequence typing for characterization of methicillin-resistant and methicillin-susceptible clones of *Staphylococcus aureus*. J Clin Microbiol 2000;38:1008-15.

7. McDougal LK, Steward CD, Killgore GE, Chaitram JM, McAllister SK, Tenover FC. Pulsed-field gel electrophoresis typing of oxacillin-resistant *Staphylococcus aureus* isolates from the United States: establishing a national database. J Clin Microbiol 2003;41:5113-20.